# Comment on "Quantum paraelectric glass state in SrCu$_3$Ti$_4$O$_{12}$" [Appl. Phys. Lett. 104, 262905 (2014)]

Shivani Sharma, Aga Shahee and N.P.Lalla[a]

UGC-DAE Consortium for Scientific Research, University campus, Khandwa road Indore, India- 452001

a) Author to whom correspondence should be addressed. Electronic mail:nplallaiuc82@gmail.com; Tel.: +91 731 2463913

We would like to comment that the claim of observing "quantum paraelectricity" (QP) and thereby emerging "quantum paraelectric glass" (QPG) state in SrCu$_3$Ti$_4$O$_{12}$ by Kumar et al, as reported in Ref. [1], is not correct. Their interpretation/conclusion [1] appears to be a result of their basic misconception about the effect of quantum fluctuations (QF) on dielectric permittivity $\varepsilon'(T)$. Due to the absence of "finger-print" of QP in their $\varepsilon'(T)$-T data, i.e. saturation of $\varepsilon'(T)$ while approaching absolute zero [2, 3], the authors have tried to prove the presence of QP using the pre-saturation region (as they have presumed) of $\varepsilon'(T)$-T. But the $\{\varepsilon'(T) -1\}^{-1}$-T and $(\varepsilon'_B/\varepsilon'_{cw} -1)$-T plots shown in Fig.4 of Ref. [1] don't even qualitatively qualify the criteria of QP. On the contrary it goes against the basic concepts of QP [2-5]. Irrespective of the type of interaction [ferroelectric (FE) or antiferroelectric (AFE)] below a temperature $T_1$ (such that $kT_1 \leq \hbar\omega_c$) the QF (with $E_c = \hbar\omega_c$) will invariably suppress the $\varepsilon'(T)$ below its CW value and will gradually saturate it at further lower temperatures ($kT << \hbar\omega_c$). Therefore in the QF regime the metric $(\varepsilon'_B/\varepsilon'_{cw} -1)$ characterizing the QP, as defined in Ref. [1], will be negative and the $\{\varepsilon'(T) -1\}^{-1}$-T will lie above the CW line. But the sign of $(\varepsilon'_B/\varepsilon'_{cw} -1)$ as reported by authors [1] is positive. The above discussed features of QP are vividly shown in Fig.1 by simulating the $\{\varepsilon'(T)-1\}^{-1}$-T and $(\varepsilon'_B/\varepsilon'_{cw} -1)$-T plots using the Barrett's formula [4] given below

$$\varepsilon'(T) = A + \frac{C}{\left(\frac{T_1}{2}\right)\coth\left(\frac{T_1}{2T}\right) - T_c} \quad \ldots\ldots (1)$$

The simulated curves in Fig.1 clearly show that the effect of AFE interaction is simply to decrease the value of ε'(T) [i.e. to shift the {ε'(T) -1}$^{-1}$-T plot vertically up], not to invert the effect of QF, as projected in Ref.1. The values of parameters used for simulation are given in Fig.1. For T>>$T_1$, the term {($T_1$/2)coth($T_1$/2T)} in (1) asymptotically approaches T [4]. Therefore, the Barrett's formula (1) will follow CW (simple linear CW when A=0 and modified CW when A>0) behaviour given by ε'(T)=A+(C/(T-$T_c$)) [3] and thus will give the same $T_c$ as that obtained by the CW fit. On the contrary the authors report two different values of $T_c$, -64K and -419K for the same ε'(T) data. Different values of $T_c$ obtained by authors is the result of two different models used in their analysis, Barrett fit (1) with A~44.54 in low-temperature (LT) regime and simple CW fit (A=0) in high-T regime. This mixed analysis is unphysical. After careful examination of the magnified view of Fig.4 of Ref. [1] one can find that the Barrett fit crosses the experimental ε'(T) data twice (at 160K & 270K) and significantly deviates from the observed data at all temperatures, more strongly beyond 300K. This is because the high-T data (>200K), as authors have themselves fitted, corresponds basically to a linear CW (A=0, $T_c$= -419K), not to a modified CW (A>0). Therefore, if QF are at all present, the (ε'-1)$^{-1}$-T should look qualitatively like the curve (a) of Fig.1 (AFE case). But it is not the case here.

We would further point out that Ferrarelli et al. [6] have shown the occurrence of QP in a related material $Na_{0.5}Bi_{0.5}Cu_3Ti_4O_{12}$ (NBCTO) through ε'(T) saturation at LT and also through temperature dependence of phonon frequency measurement using LT IR-reflectivity. It seems that authors of Ref.1 have taken the report of Ferrarelli et al [6] as granted and keeping in view only the similarity between the crystal structures, have presumed the presence of QP in their SCTO sample too. It should be noted that QP behavior is very sensitive to change in composition [2]. Since in SCTO the A-site occupancy is totally different from that of NBCTO, therefore QP seen in NBCTO [6], may not necessarily be present in SCTO also. For QP phonon softening on cooling [5] is required. Authors refer to the LT Raman work by Mishra et al. [7] as the only support for the occurrence of phonon softening in SCTO and try to correlate the observed drop in the $\varepsilon'(T)$ to the phonon softening. This correlation is against the established theory. According to Eq.1 of Kvyatkovskiĭ [5] phonon-softening will cause an increase of $\varepsilon'(T)$ on cooling [2] and not a decrease. A drop in $\varepsilon'(T)$ will be caused by phonon hardening. If the spin-phonon coupling is inducing phonon hardening below $T_N$, decrease of permittivity below $T_N$ should be frequency independent as in $EuTiO_3$ [8]. On the contrary, SCTO exhibits strong frequency dispersion below $T_N$ [1]. It means, observed decrease of $\varepsilon'(T)$ below $T_N$ cannot be attributed to phonon hardening also. Authors attribute the frequency dispersion below $T_N$, as shown in Fig.5 (a) of Ref. [1], to the "underlying QF". If QF is still "underlying" then saturation of $\varepsilon'(T)$ is inevitable but this essential signature of the presence of QF has not been seen in the data. Taking $T_1$=155K [1] the estimated value of the QF frequency is ~3.1THz. The experimentally measured value of the soft phonon frequency at ~30K as given in Ref.[7] is ~13.3 THz (~443.1 $cm^{-1}$). How come the presumed "underlying QF" can cause QP as its frequency (~3.1 THz) is more than 4 times slower than the experimentally measured soft phonon frequency (~13.3 THz). Hence the description of QP given in ref [1] is physically inexplicable. Please note that in NBCTO [6], the measured soft phonon frequency is ~23$cm^{-1}$ (0.69 THz) which is more than 2 times slower than the QF frequency of ~ 1.7 THz ($T_1$=85K). It is thus clear that the proposition of

QP in SCTO in Ref. [1] is totally unphysical. The observed strong frequency dispersion below $T_N$ is a clear signature of dielectric relaxation with a broad distribution of relaxation frequencies [9]. In such case the "incipient ferroelectric-like" increase of permittivity seen on cooling is also not caused by phonon softening but by some soft dielectric relaxation.

Thus based on the above described simulations and physical arguments given against the analysis in Ref. [1], we conclude that the report of "quantum paraelectricity" and thereby emerging QPG state in $SrCu_3Ti_4O_{12}$, as claimed in Ref. [1] is not correct.

**Figure captions:**

**Figure 1.** $(\varepsilon'-1)^{-1}$-T and $(\varepsilon'_B/\varepsilon'_{cw}-1)$-T plots simulated based on Barrett formula for QPs (a) for linear CW i.e. A=0 and (b) for modified CW with A=45 for paraelectric materials with AFE and FE interactions.

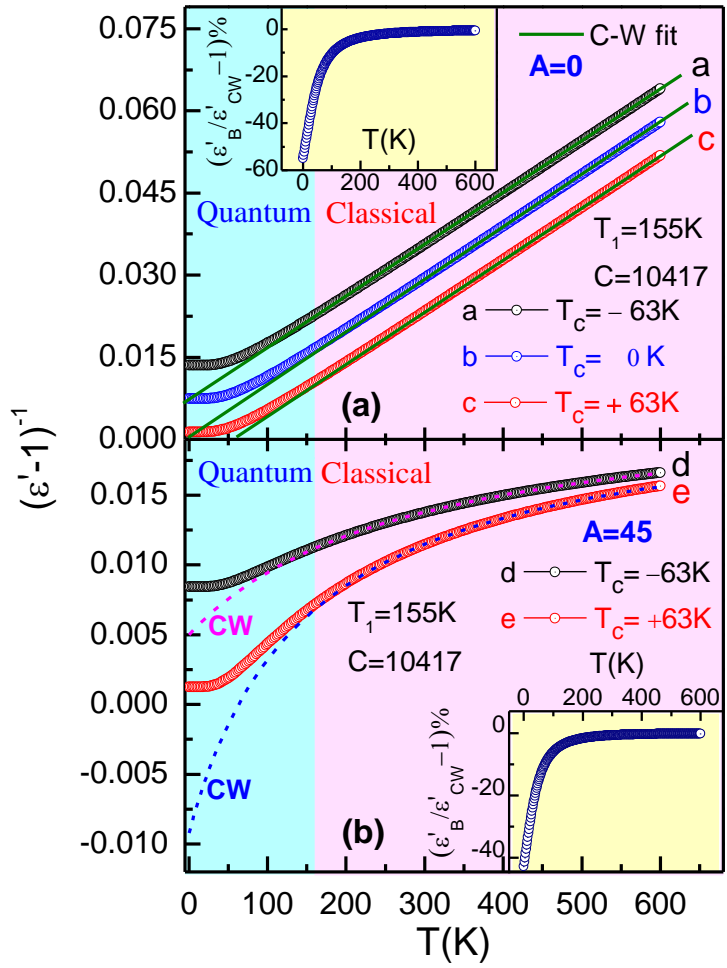

Figure 1